\documentclass[sigconf,10pt,nonacm]{acmart}

\usepackage{booktabs}
\usepackage{graphicx}
\usepackage{subcaption}
\usepackage{amsmath}

\usepackage{xspace}
\usepackage{balance}

\graphicspath{{figs/}}

\setcopyright{rightsretained}
\acmConference[HotMobile '27]{The 28th International Workshop on Mobile Computing Systems and Applications}{TBD 2027}{TBD}
\acmYear{2027}
\copyrightyear{2027}
\AtBeginEnvironment{thebibliography}{\fontsize{9}{10}\selectfont}
\AtBeginDocument{\setlength{\bibsep}{0pt}}

\setcopyright{none}
\begin{document}

\title{What Is the Cost of a Handover Interruption? Rethinking 5G Mobility for Aerial Vehicles}

\author{André Melo}
\affiliation{%
  \institution{Aeronautics Institute of Technology, Brazil}
   \country{Brazil}
}
\email{andre.melo@ita.br}

\author{Yueqi Chen}
\affiliation{%
  \institution{Northwestern University, USA}
   \country{USA}
}
\email{yueqichen@northwestern.edu}

\author{Igor Kadota}
\affiliation{%
  \institution{Northwestern University, USA}
   \country{USA}
}
\email{kadota@northwestern.edu}

\author{Lourenço Alves Pereira Jr.}
\affiliation{%
  \institution{Aeronautics Institute of Technology, Brazil}
   \country{Brazil}
}
\email{lourenco.junior@gp.ita.br}

\begin{abstract}
Low-altitude aviation is moving toward large-scale operations increasingly dependent on 5G cellular connectivity. However, 5G mobility remains blind to the operational risk of its decisions:
the same interruption may be harmless in cruise over an empty field but critical near obstacles, crowds or other aircraft. Although 5G already provides mechanisms to obtain location, population density, and radio conditions, mobility decisions do not translate this information into an operational risk metric.
We propose Risk-Aware Mobility (RAM), which maps predicted communication disruptions to an operational-risk proxy using
information staleness, position uncertainty, and operational context, and exposes this risk to existing mobility mechanisms. In a dense S\~ao Paulo deployment with 188 licensed 5G sectors, risk-aware handover timing reduces the mean
operational-risk proxy by 45\% relative to conventional A3 without degrading serving-link quality. These results motivate shifting 5G mobility from optimizing handover performance to managing the operational risk of mobility decisions. Beyond handover timing, this perspective raises a broader question: which cells should serve aerial operations
when connectivity is evaluated by operational risk rather than radio performance alone?

\end{abstract}



\maketitle

\section{Introduction}
\label{sec:intro}

Low-altitude airspace is undergoing a fundamental transformation.
Aviation authorities worldwide are developing traffic-management
frameworks for large-scale operations of unmanned and increasingly
autonomous aircraft \cite{USDOT_AAM_National_Strategy_2025,
sesarConops,dca351-6,uk_conops,uk_roadmap}.
The United States has established a national strategy for Advanced Air
Mobility, while the United Kingdom targets routine beyond-visual-line-of-sight
(BVLOS) drone operations by 2027. Commercial UAV delivery services
are already emerging in China, the Middle East, and
Brazil \cite{meituan,keeta_dubai,anac_speedbird}.
Scaling these operations requires timely exchange of aircraft position,
telemetry, and traffic information among aircraft,
operators, and airspace-management services \cite{faaConops,
sesarConops,dca351-6}.

Commercial cellular networks are a promising component of this
infrastructure, offering wide-area coverage, high-capacity communication,
and integration with edge services \cite{USDOT_AAM_National_Strategy_2025,
USDOT_AAM_Comprehensive_Plan_2025}.
The question is shifting from whether 5G can connect aerial vehicles
to whether its mechanisms can support their operational requirements.

Unlike terrestrial users, UAVs experience aerial propagation that exposes
them to multiple base stations, irregular antenna sidelobes, and rapid
changes in serving-cell quality \cite{tr36777,euler2019mobility,
huang2024sky}.
\textbf{This creates a fundamental mobility trade-off: switching cells
may interrupt communication, while staying connected may lead to
radio-link failure.}
Either outcome can interrupt delivery of fresh aircraft state.
Commercial measurements confirm frequent aerial handovers and
second-scale data-plane outages even after nominally successful
transitions \cite{liuLowAltitude2026}.

Equal interruptions need not imply equal operational risk.
Consider two UAVs experiencing the same 500-ms data-plane gap: one over
sparsely populated ground, the other near a crowd or another aircraft.
As position information becomes stale, uncertainty about the aircraft's
location grows; its consequences depend on surrounding traffic,
operating constraints, and ground exposure.
Consistent with this operation-dependent view, the Specific Operations
Risk Assessment (SORA) requires communication performance appropriate
to the intended operation \cite{sora2024,jarus_sora25_annexf}.

Existing mobility mechanisms are designed to maintain connectivity
and limit service disruption. Mobility Robustness Optimization (MRO)
addresses too-early, too-late, and wrong-cell handovers; Conditional
Handover (CHO) prepares targets in advance; and Dual Active Protocol
Stack (DAPS) can reduce service interruption \cite{ts38300r16,
ts38300r18}. UAV-specific approaches further exploit flight paths
and radio conditions to improve mobility decisions \cite{bai2019route,
azari2020ml}.
These mechanisms serve their intended purposes, but were not designed
to distinguish the operational consequences of the same communication
disruption across different aerial contexts.

\textbf{The gap is not how 5G executes mobility, but what its decisions
optimize.}
Radio measurements can guide when to switch or stay, but do not express
what losing fresh aircraft information would mean at that point of
the flight. Rather than replacing existing mobility procedures,
we ask whether risk can become an additional decision
metric. This gap extends beyond handover timing.
Radio performance also determines which cells
are preferred along an aerial route, without
necessarily reflecting the operational consequences
of their selection.

We propose \emph{Risk-Aware Mobility} (RAM), which maps
communication induced information staleness into position uncertainty
and an aircraft-specific risk proxy.
RAM uses this quantity to inform existing mobility decisions,
without requiring new air-interface procedures or UE modifications.
Its inputs can be derived from network measurements, 5G positioning
and analytics functions, and geographic exposure data \cite{ts23273,
ts23288}; 3GPP aerial-UE flight-path reporting could further refine
the estimate \cite{ts36331r15,ts38300r18,ts38331r18}.

\textbf{Can an operational-risk proxy improve 5G mobility decisions?}
We investigate this question with commercial 5G traces, OpenAirInterface
(OAI) experiments, and a dense S\~ao Paulo deployment reconstructed
from 188 licensed 5G sectors \cite{anatelMosaico,tr38901,tr36777}.
Our evaluation reveals a non-monotonic trade-off: suppressing
handovers initially lowers operational risk, but excessive deferral
increases it as the serving link deteriorates.
Risk-aware handover timing reduces the mean risk proxy
from 0.289 to 0.160 (45\%) relative to a conventional 3-dB A3
configuration while maintaining serving-link quality.
This is a reduction in a modeled risk proxy, not a
measured reduction in accident probability.

Our contributions are threefold:
\textbf{(i)} We characterize the competing costs of switching and
staying using commercial 5G traces and OAI experiments.
\textbf{(ii)} We introduce RAM, which extends
an existing operational-risk abstraction to
5G mobility decisions \cite{melo2026riskaware}, enabling risk-aware
handover timing. 
\textbf{(iii)} We show that risk-aware handover timing lowers modeled
operational risk beyond conventional A3 tuning in a dense urban
deployment.

\section{When should a UAV switch cells, and when should it stay?}
\label{sec:motivation}

A mobility controller must balance the immediate disruption of switching cells against the increasing risk of staying on a deteriorating link. We characterize this trade-off using commercial 5G traces and controlled OAI experiments.

\subsection{Observation 1: Switching Can Interrupt Data Delivery}
\label{sec:motivation:dataset}

We analyze 547 traces from the 5G SA measurements of Liu
et al.\cite{liuLowAltitude2026}, spanning four operators, several
altitudes, and rural and urban environments.
We identify a \emph{post-handover data-plane outage} when a handover
recorded as successful is followed by at least one second without
L1 throughput within a two-second observation window.
This trace-derived criterion is not itself a 3GPP handover-failure
classification.

\textbf{Successful handover does not guarantee continuous data delivery.}
At 300\,m, aerial UEs experience 14.6 handovers/km, compared with
about 9.2/km on the ground.
Of these aerial transitions, 16.9\% are followed by a post-handover
data-plane outage, with a median duration of about 2.2\,s.
Along the urban--rural route, the measured fraction varies from
0.14 in its rural segment to 0.22 near the urban--rural transition.
Its duration, however, does not depend on where it happens: the
median is 1.9--2.2\,s at every altitude from 80 to 300\,m. What varies
along a flight is how likely the outage is and what lies beneath.

Liu et al. identify a related mechanism, termed
\emph{pseudo-successful handover}, in which control-plane signaling
succeeds but target uplink access fails \cite{liuLowAltitude2026}.
The trace-derived outages can include events consistent with this
mechanism, but do not individually establish their cause.

These measurements reveal a cost of switching that conventional
handover-success indicators may not capture: a completed mobility
procedure can still leave the aircraft temporarily unable to deliver
fresh information.
However, avoiding handovers altogether is not a solution, as the
serving link may deteriorate while the aircraft continues moving.

\subsection{Observation 2: Staying Can Destroy Connectivity}
\label{sec:motivation:oai}

We investigate the competing costs of switching and staying in a
controlled OAI testbed \cite{oai}, with one CU, two DUs on different
carriers, and one UE in the RF simulator.
Per-link attenuation is varied along a flight trajectory while a
UDP probe measures data-plane continuity.

\textbf{Switching has an immediate cost.}
A regular inter-frequency handover introduces a 377-ms data-plane
gap in our setup.
Deliberately blocking target uplink access causes repeated
random-access failures and a multi-second interruption.
These results illustrate why postponing a handover may initially
appear beneficial for continuous data delivery.

\textbf{Staying has an increasing cost.}
The serving link withstands approximately 23\,dB of additional
attenuation before deteriorating sharply, with radio-link failure
between 30 and 35\,dB.
In one controlled run, excessive handover deferral lost 43\% of probe
packets, compared with 1.9\% when deferral was constrained by the
serving-link limit.
These are results for the tested configuration, not universal 5G
thresholds.

Thus, postponing a handover can avoid an immediate interruption,
but only while the serving link remains capable of carrying data.
Beyond that point, the accumulated cost of staying can exceed the
cost of switching.

\noindent\textbf{Takeaway.}
Switching and staying impose competing communication costs, but radio measurements alone cannot determine their operational consequences. Comparing these alternatives requires a risk metric that accounts for the aircraft and its operating context.

\section{From Communication Disruption to Operational Risk}
\label{sec:design}

We now map communication-induced information
staleness into an operational-risk proxy for
comparing candidate handover decisions.

\noindent\textbf{From staleness to uncertainty.}
Consider a UAV transmitting aircraft status and,
when required by its operation, video for remote
situational awareness. Let $\Delta_s(t)$ denote
the age of the latest status update received by
the network. During an interruption, the network
continues relying on that update even as the UAV
moves.

Following the conservative containment abstraction
used in\cite{sora2024,jarus_sora25_annexf},
we bound its position uncertainty by
\begin{equation}
r(t)=r_{\min}+v_{\max}\Delta_s(t),
\label{eq:uncertainty}
\end{equation}
where $r_{\min}$ is baseline localization uncertainty
and $v_{\max}$ bounds aircraft speed. The corresponding
region $\mathcal B(t)$ is a disk centered at the
last reported position. This is a bound on the
network's knowledge of the aircraft, not an estimate
of navigation error or loss of onboard control.

\noindent\textbf{From uncertainty to consequence.}
We retain three operational consequences \cite{melo2026riskaware} from
the risk abstraction in \cite{sora2024,jarus_sora25_annexf}:
consumed separation margin, ground exposure, and
loss of video coherence.

Let $d_h(t)$ denote the distance from the last
reported position to each relevant hazard $h$
(another aircraft, a restricted volume, an obstacle)
and let $r_h(t)$ include its protection distance and
positional uncertainty. We define
\begin{align}
R^N(t)
&=\frac{1}{\bar R^N}\sum_{h}[r(t)+r_h(t)-d_h(t)]^+,
\label{eq:nav-risk}\\
R^G(t)
&=\frac{1}{\bar R^G}
\int_{\mathcal B(t)}
\rho(\mathbf x,t)\,d\mathbf x,
\label{eq:ground-risk}\\
R^V(t)
&=\frac{[z(t)-H_{\mathrm{nom}}]^+}{\bar R^V}.
\label{eq:video-risk}
\end{align}
Here, $\rho(\mathbf x,t)$ is population density,
$[x]^+=\max\{x,0\}$, and
$z(t)=\sum_{i=1}^{m}\alpha_i\Delta_v^{(i)}(t)$,
with $\sum_i\alpha_i=1$, is the weighted age of the
$m$ most recently received video frames, where
$\Delta_v^{(i)}(t)$ is the age of the $i$-th most
recent frame.

\begin{figure}[t]
    \includegraphics[width=0.8\columnwidth]{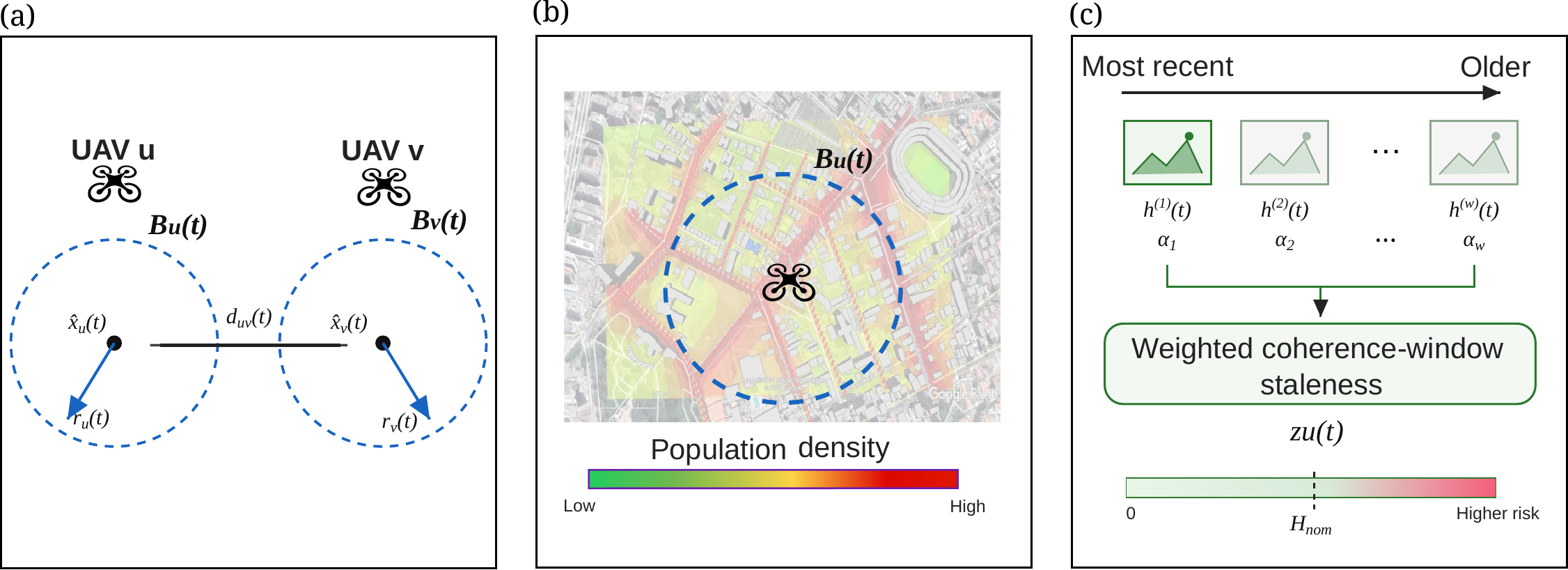}
    \vspace{-0.5cm}
    \caption{The same staleness can induce different operational consequences. (A) Navigation risk increases as position uncertainty consumes the separation margin between neighboring UAVs. (B) Ground risk depends on the population exposed within the UAV's uncertainty region, so the same AoI can imply different risk across locations. (C) Video-coherence risk captures the weighted staleness of the most recent completed frames and increases once it exceeds the nominal coherence threshold. The three components are normalized at their corresponding operational boundaries.
}
    \label{fig:risk-components}
    \vspace{-0.5cm}
\end{figure}

The navigation component measures a separation
deficit, while the ground component measures
population exposure within the uncertainty region;
neither is a collision or casualty probability.
The video measures excess staleness
relative to a task-specific coherence requirement.
It is relevant only when timely video supports
the intended operation.

\noindent\textbf{Operational reference levels.}
The three components have different physical units.
Following the normalization approach used for
risk-aware UAV communications, we scale them by
positive reference levels derived from declared
operational limits. For example, the ground
reference can be defined as
$\bar R^G=\rho_{\max}\pi r_\tau^2$, where
$r_\tau=r_{\min}+v_{\max}\tau_s^{\max}$
and $\tau_s^{\max}$ is the declared maximum
status-update age. Similarly,
$\bar R^V=\tau_v^{\max}-H_{\mathrm{nom}}$
defines the video reference at its maximum
tolerable coherence age. The navigation reference
is determined by the declared separation margin.

The resulting composite proxy is
\begin{equation}
\boxed{
R(t)=w_N R^N(t)+w_G R^G(t)+w_V R^V(t)
}
\label{eq:risk}
\end{equation}
where $w_N+w_G+w_V=1$, with nonnegative weights.
Reference levels and weights are fixed when
comparing mobility decisions. The level $R=1$
defines a declared composite reference boundary,
not a certified safety threshold. The metric
quantifies communication-induced operational
consequences rather than accident probability.

\noindent\textbf{From risk to handover timing.}
A handover affects operational risk through the
freshness of the information available to the
network. The interruption duration depends on
whether the transition succeeds normally,
experiences a prolonged data-plane outage,
or is preceded by failure of the serving link.

For a candidate execution time $\tau$, let
$R(s\mid\tau)$ denote the operational-risk
trajectory resulting from these possible
communication outcomes. We define
\begin{equation}
C(\tau)=
\mathbb E\left[
\int_t^{t+H}R(s\mid\tau)\,ds
\right],
\label{eq:mobility-cost}
\end{equation}
where $H$ is a common prediction horizon and
the expectation is taken over possible outage
durations and link-failure outcomes.
Candidate decisions are evaluated using the
same operational context and outage model.

RAM selects
\begin{equation}
\tau^*=
\arg\min_{\tau\in\mathcal T_{\mathrm{feasible}}}
C(\tau),
\label{eq:ram-decision}
\end{equation}
where $\mathcal T_{\mathrm{feasible}}$ contains
execution times satisfying the existing radio
conditions and serving-link viability constraints.

RAM does not change the handover procedure or
select a different target cell. It retains the
A3-selected target and uses operational risk
to determine when the transition should occur.
The required context may be supplied by aviation
services and combined with network measurements,
positioning, and analytics
\cite{ts23273,ts23288}.
Evaluating Eq.~\eqref{eq:mobility-cost} requires
predicting where the aircraft will be and what
radio conditions it will see; Section~\ref{sec:case}
compares a predictor that knows the trajectory with one that uses only what the
network observes.

\section{Evaluation}
\subsection{Does Risk Awareness Improve Mobility Decisions?}
\label{sec:case}

\begin{figure}[t]
  \centering
  \includegraphics[width=\columnwidth,height=0.60\textheight,keepaspectratio]{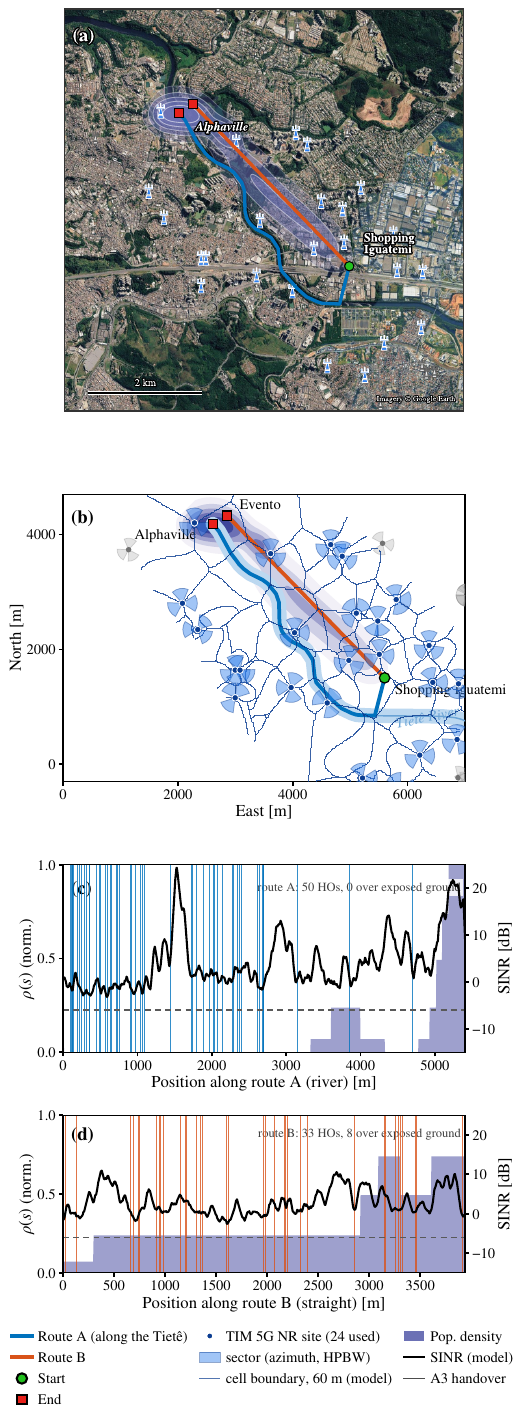}
  \caption{The Alphaville delivery corridor.
(a)~Geographic overview showing the two routes,
licensed TIM 5G NR sites, and ground-exposure levels.
(b)~Local view of the licensed sectors, including
antenna azimuths, beamwidths, and best-server cell
boundaries at 60\,m predicted by the channel model.
(c)--(d)~Ground exposure and predicted SINR along
the river route~A and straight route~B, with
serving-cell transitions under a 3-dB A3 trigger.}
  \label{fig:scenario}
  \vspace{-12pt}
\end{figure}

\textbf{Experimental setting.}
We reconstruct a UAV corridor around Alphaville,
S\~ao Paulo, using licensed 5G NR deployments from
Anatel's Mosaico database \cite{anatelMosaico}.
As shown in Fig.~\ref{fig:scenario}(a)--(b),
the deployment contains 59 base stations and
188 sectors operating at 3.55 GHz within 5 km
of the corridor; the simulation uses the 72 sectors
within 1.5\,km of the routes (59 after pruning those
never within 20\,dB of the best server).

We model aerial propagation using 3GPP urban-macro
path loss, aerial line-of-sight probability,
antenna azimuth, tilt, gain, correlated shadowing,
and inter-cell interference \cite{tr38901,tr36777}.
Route~A follows the Tiet\^e river for 5.4\,km and
exposes little population; route~B is the 3.9-km
straight line from the mall and crosses residential
areas. Six UAVs fly both routes in both directions at
60\,m and 14\,m/s, with departures spread over
300\,s, for three seeds. Ground exposure $\rho$ is
the population level set of the corridor;
``exposed ground'' denotes normalized exposure
above 0.3.

We compare conventional A3 with hysteresis
values of 3, 6, 9, and 12 dB against RAM.
All policies experience the same channel
realizations and handover-outage outcomes.
Outage probabilities and durations are
derived from the commercial 5G traces
examined in Section~\ref{sec:motivation}:
a regular handover interrupts delivery for
21.4\,ms and, with probability 0.081, becomes
a post-handover outage drawn from the 133
episodes measured at 80\,m (median 2.2\,s).
The proxy uses equal weights, $v_{\max}=14$\,m/s,
the other aircraft as hazards and
$\bar R^V=H_{\mathrm{nom}}$; RAM uses a 30-s
horizon on a 1-s grid.

RAM retains the A3-selected target cell,
but shifts handover execution within a
short prediction horizon to reduce expected
operational risk while preserving
serving-link viability.
We evaluate two predictors of $R(s\mid\tau)$.
\emph{RAM-T} (trajectory) uses the planned path
and the simulated future channel: an upper bound
on what context can deliver.
\emph{RAM-R} (reachability) uses only what the
network observes: the last reported position and
heading, exposure averaged over a $20^\circ$ cone
of reachable positions, and a shadow-free radio
map corrected by the current measurement with a
5-s decay.

\begin{figure}[t]
  \centering
  \includegraphics[width=0.62\columnwidth]{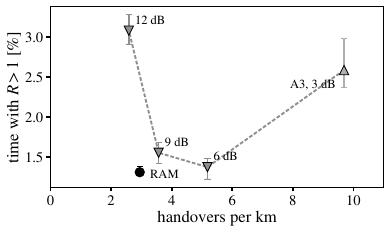}
  \caption{Time above the risk boundary vs.\ handovers
per km. Dashed: the A3 hysteresis sweep (3--12\,dB); black
circle: RAM-T; diamond: RAM-R. Fewer handovers first reduce
and then increase risk. Error bars span the three seeds.}
  \label{fig:frontier}
  \vspace{-6pt}
\end{figure}

\textbf{Finding 1: Fewer handovers are not
necessarily safer.}
Table~\ref{tab:case} and Fig.~\ref{fig:frontier}
reveal a non-monotonic relationship between
handover frequency and operational risk.

Increasing A3 hysteresis from 3 to 9 dB
reduces handovers from 9.7 to 3.6 per km
and lowers mean operational-risk proxy
from 0.289 to 0.171. However, increasing
hysteresis further to 12 dB raises risk
to 0.221, despite reducing handovers
to only 2.6 per km.

The reason is the mobility trade-off
identified in Section~\ref{sec:motivation}:
suppressing unnecessary transitions
reduces interruption exposure, but holding
the serving connection too long allows
radio quality to deteriorate.
At 12 dB, average serving SNR falls to
1.9 dB, and time above the declared
risk boundary increases to 3.1\%.

\begin{table}[t]
\centering
\footnotesize
\setlength{\tabcolsep}{3.5pt}
\caption{Mobility decisions in the S\~ao Paulo
deployment. $R>1$ denotes time above the
declared operational-risk boundary.
Results use three simulation seeds.}
\label{tab:case}
\begin{tabular}{lrrrr}
\toprule
Policy & HO/km & $\bar R$ & $R>1$ & SNR \\
\midrule
A3, 3 dB  & 9.7 & 0.289 & 2.6\% & 3.8 \\
A3, 6 dB  & 5.2 & 0.178 & 1.4\% & 3.4 \\
A3, 9 dB  & 3.6 & 0.171 & 1.6\% & 2.7 \\
A3, 12 dB & 2.6 & 0.221 & 3.1\% & 1.9 \\
RAM-T     & 2.9 & \textbf{0.160}
          & \textbf{1.3\%} & 3.4 \\
RAM-R     & 3.2 & 0.198 & 1.4\% & 3.3 \\
\bottomrule
\end{tabular}
\end{table}

\textbf{Finding 2: Risk awareness changes
which handovers are executed.}
RAM-T achieves the lowest mean operational-risk
proxy (0.160), a 45\% reduction relative
to A3 at 3 dB, in each of the
three seeds and not only on average, while
keeping the serving SNR within 0.4\,dB of
the 3-dB trigger.

More importantly, the improvement does not
come simply from suppressing handovers.
RAM-T performs 2.9 handovers/km, close to
the 12-dB A3 configuration, but achieves
lower risk and substantially better
serving-link quality.

The difference is which transitions are
executed. Along the crowd-exposed route~B,
conventional A3 at 3 dB executes 94 handovers
over exposed ground per run, compared with
30 under RAM-T; the associated data-plane
mute time over exposed ground decreases
from approximately 23 s to 12 s.
These are handovers RAM-T drops, not moves:
the share of transitions over exposed ground
is 49\% for both policies, and no executed
handover is shifted by more than 0.4\,s.
Hysteresis suppresses transitions by signal
margin; RAM drops those whose predicted
advantage does not outlast the horizon,
including all 110 ping-pong transitions per
run of the 3-dB trigger, and keeps the rest
at the serving quality of the 3-dB trigger.

\textbf{Finding 3: The gain does not require
knowing the trajectory.}
RAM-R reproduces the mechanism with network
information only: 3.2 handovers/km, no
ping-pong, the same 2.7 outages over exposed
ground per run as RAM-T, 1.4\% of time above
the boundary against 1.3\%, and serving SNR
within 0.1\,dB. Its mean proxy (0.198) differs
from RAM-T only in one seed, where a single
long outage shared by all policies falls
worse for RAM-R; the other two seeds coincide.
Not knowing the trajectory costs little here.

\textbf{What does this tell us?}
Conventional mobility optimization already
balances the radio costs of switching
and staying. Our results suggest that
operational risk provides an additional
dimension for navigating that trade-off.
Rather than making every handover faster
or simply reducing their number, RAM
prioritizes when interruptions can be
better tolerated by the aerial operation.

\section{Discussion and Open Questions}
\label{sec:discussion}

\textbf{What should mobility optimize?}
Existing mobility mechanisms are designed to
maintain connectivity and reduce service disruption.
Our results do not challenge these objectives.
Instead, they expose a limitation of decisions
based only on radio performance: two handovers
with similar communication costs may create
different levels of operational risk.
Adjusting A3 hysteresis changes how frequently
handovers occur, but cannot explicitly distinguish
where an interruption is safer to tolerate.
RAM introduces operational risk as an additional
decision metric, without changing how handovers
are executed.

\textbf{Can a more intelligent RAN become risk-aware?}
Advances in 5G positioning, integrated sensing
and communication (ISAC), and network analytics
are expanding the information available to
cellular networks. Combined with flight information
and airspace-management services, these capabilities
could allow the RAN to estimate the operational
context of individual aircraft.
However, knowing more about the environment does
not necessarily lead to safer mobility decisions.
The network still needs a way to translate this
information into a metric its mobility mechanisms
can use.

RAM provides such a connection by mapping
information staleness, position uncertainty,
and operational context into a common risk proxy.
This suggests a path toward risk-aware mobility
in which existing radio measurements determine
feasible transitions, while operational risk
informs their timing.
Such decisions could be integrated into radio
resource management or supported by an external
controller, without introducing new handover
procedures. How to expose and update this risk
information within the RAN remains open.

\textbf{What are the limits of risk awareness?}
Our metric is an operational-risk proxy, not
an estimate of accident probability.
Its weights and reference levels depend on the
operation, and its accuracy depends on the
freshness and quality of the available context.
Likewise, our evaluation uses simulated radio
conditions and commercial trace-derived outages
in one urban deployment. These results show
that risk awareness can improve decisions under
the modeled assumptions, but do not establish
a measured improvement in flight safety or
performance in a commercial RAN.
In this corridor the exposure varies little
within the 30-s horizon, so RAM changes which
handovers happen rather than where; moving
them requires an exposure gradient the
aircraft can reach in time.

\textbf{Who should define acceptable risk?}
Cellular operators can characterize radio
reliability, but aviation services and aircraft
operators define the safety requirements of
their operations. Should the RAN independently
estimate operational risk, or should aviation
services expose risk limits that mobility
mechanisms can directly use? How should the
network react when this information is uncertain,
unavailable, or changes during flight?

These questions extend beyond aerial vehicles
to other autonomous systems that depend on
cellular connectivity. As networks become
increasingly capable of sensing and understanding
their environment, mobility decisions may need
to evolve accordingly. The question is no longer
only when a handover can succeed, but whether
the network can account for the operational
risk of executing it.

\textbf{Can operational risk inform which cells should serve UAVs?} Risk awareness need not be limited to deciding when a handover occurs. The same operational-risk metric could inform which cells are eligible to serve aerial vehicles along a corridor or within a geographic region. A cell offering stronger radio coverage may still lead to frequent interruptions in operationally sensitive areas, while a different serving-cell configuration may provide lower overall risk. This raises a broader question: should aerial connectivity be planned solely around radio coverage, or should network operators and aviation services jointly identify serving-cell configurations that satisfy operational requirements? Such an approach could extend risk awareness from individual mobility decisions to aerial service planning, without requiring changes to the underlying handover procedures.

\balance
\bibliographystyle{ACM-Reference-Format}
\bibliography{references}

\end{document}